\documentclass[english,onecolumn]{IEEEtran}
\usepackage[T1]{fontenc}
\usepackage[letterpaper]{geometry}
\usepackage{float}
\usepackage{amsmath}
\usepackage{amsthm}
\usepackage{amssymb}
\usepackage{graphicx}
\usepackage{cite}

\makeatletter

\providecommand{\tabularnewline}{\\}

\newtheoremstyle{rmd}
  {}
  {}
  {\em}
  {}
  {\em}
  {}
  {5pt}
  {\thmname{#1}\thmnumber{ #2}.\thmnote{ #3}}
\newtheorem{theorem}{Theorem}
\newtheorem{assumption}{Assumption}
\newtheorem{alg}{Algorithm}
\newtheorem{thm}{\protect\theoremname}
\theoremstyle{remark}
\newtheorem{rem}[thm]{\protect\remarkname}

\newcommand{\dt}{\frac{d}{dt}}

\makeatother

\usepackage{babel}
\providecommand{\remarkname}{Remark}
\providecommand{\theoremname}{Theorem}

\begin{document}
\title{Multivariable Super-Twisting Algorithm for Systems with Uncertain Input Matrix and Perturbations$^\star$} 

\author{\vspace*{5mm}Jaime A. Moreno$^{\dagger}$, H\'{e}ctor R\'{i}os$^{\ddagger*}$, Luis Ovalle$^{\mathsection}$ and Leonid Fridman$^{\mathsection}$
	\thanks{$\dagger$Universidad Nacional Aut\'{o}noma de M\'{e}xico (UNAM), Instituto de Ingenier\'{i}a, 04510, Mexico City, Mexico. Email: {\tt\small JMorenoP@ii.unam.mx}}
	\thanks{$^{\ddagger}$Tecnológico Nacional de México/I.T. La Laguna, División de Estudios de Posgrado e Investigación, Blvd. Revolución y Cuauhtémoc S/N, C.P. 27000, Torreón, Coahuila, Mexico.}
	\thanks{$^{\mathsection}$Universidad Nacional Aut\'{o}noma de M\'{e}xico (UNAM), Facultad de Ingenier\'{i}a, 04510, Mexico City, Mexico.  }	
	\thanks{$^*$Cátedras CONACYT, Av. Insurgentes Sur 1582, C.P. 03940, Mexico City, Mexico. Email: {\tt\small hriosb@correo.itlalaguna.edu.mx}}
	\thanks{$^\star$The authors gratefully acknowledge the financial support from CONACYT project 282013, PAPIIT-UNAM projects IN115419 and IN102121, C\'{a}tedras CONACYT CVU 270504 project 922, and TECNM research projects.}
	\thanks{This paper has been submitted for possible publication to IEEE Transactions on Automatic Control.}}
\maketitle
\begin{abstract}
This paper proposes a Lyapunov approach to the design of a multivariable
generalized Super-Twisting algorithm (MGSTA), which is able to control
a system with perturbations and uncertain control matrix, both depending
on time and the system states. The presented procedure shows that, under reasonable assumptions for the uncertainties, it is always possible to find a set of constant gains for the MGSTA in order to ensure global and robust finite-time stability of the system's outputs. Simulation results on an omnidirectional mobile robot illustrate the performance of the MGSTA. 
\end{abstract}
\begin{IEEEkeywords}
Sliding-mode control, Super-Twisting algorithm, Robust control, Uncertain systems.        
\end{IEEEkeywords}

\section{Introduction}
The Super-Twisting algorithm (STA) (see \cite{levant} and \cite{LEVANT1998})
is one of the most cited non-linear controllers of the last two decades.
The STA yields finite-time stabilization for a perturbed system with
relative degree one using only output information. The STA achieves
a second-order sliding motion providing a continuous control signal.

In the first stage of development, the convergence analysis of the
STA was made based on a geometric approach (see \cite{levant} and
\cite{LEVANT1998}) considering the effect of time-varying uncertain
control gains and perturbations. It took fifteen years until the first
\textit{strict} Lyapunov function ensuring finite-time convergence
for the STA, was presented in \cite{moreno2008} (see also \cite{moreno_osorio}).
The design of the Lyapunov functions in \cite{moreno2008} and \cite{moreno_osorio},
pushed forward the analysis of the properties of STA and its applications.

Recently for the scalar case, in \cite{seeber} and \cite{seeber4},
the most general conditions for the gain selection of the STA have
been found. Different approaches to estimate the convergence time
for STA are summarized in \cite{seeber2} and \cite{seeber5}. Moreover,
in \cite{behera} some new gain conditions for the STA are provided
to guarantee the stability of an uncertain system with actuator saturation.
A modified version of the STA has been proposed by \cite{seeber3}
in order to ensure a saturated control signal. Similarly, in\cite{castillo}
and \cite{sat} some modifications to the structure of the STA are
made to ensure the saturation of the control signal. In \cite{haimovich},
a generalization of the STA has been presented to deal with a larger
class of perturbations. In \cite{ulises2} the STA gains have been
designed to minimize the amplitude of chattering or the energy needed
to maintain the real second-order sliding-mode. The previously mentioned
approaches gave way for the usage of the STA in different applications:
wind turbines \cite{evangelina}, pneumatic actuators \cite{taleb},
fuel cells \cite{kunusch}, quadrotor helicopters \cite{denis}
and \cite{rev} and many others. Applications of \cite{levant} --\cite{ulises2} clarify two main restrictions of these results: 
\begin{enumerate}
\item The perturbations are assumed to have an \textit{a priori} known Lipschitz
constant, preventing them to grow with the states. 
\item The presence of state-dependent uncertainty in the control gain is
not considered.
\end{enumerate}

These two problems have been solved in \cite{isma} adding linear
terms to the discontinuous control law. However, since the previous
results allow only for a scalar control input, a major problem remains:
only systems with multiple inputs can be taken into account if they
can be decoupled into scalar subsystems that can be treated independently.
This restricts considerably the possible applications, since uncertain
couplings between control channels cannot be considered.

To deal with multiple-input systems some multivariable generalized
Super-Twisting algorithms (MGSTA) have been developed. The Super-Twisting algorithm is a second-order sliding-mode controller and its structure is given by
\begin{equation*}
\left[\begin{array}{c}
u\\\dot{v}
\end{array}
\right]=\left[\begin{array}{c}
-k_1\phi_1(x)+v\\-k_2\phi_2(x)
\end{array}\right],
\end{equation*}
where $x\in\mathbb{R}^n$ is the variable to control, $u\in\mathbb{R}^n$ is the control signal, $v\in\mathbb{R}^n$ is an internal state of the algorithm and $k_1,\ k_2\in\mathbb{R}$ represent the algorithm gains. The functions $\phi_1$ and $\phi_2$ are considered to be of the form $\phi_1(x)=(\alpha\|x\|^{-p}+\beta)x$ and $\phi_2(x)=[\alpha(1-p)\|x\|^{-p}+\beta](\alpha\|x\|^{-p}+\beta)x.$ The first MGSTA design was proposed in \cite{edwards}, based on the Lyapunov function suggested in \cite{moreno_osorio}, with a Lipschitz perturbation.
However, the presence of an uncertain input matrix is not considered.
Two kinds of MGSTAs were introduced in \cite{LopMor15} and \cite{moreno}, including
the one proposed by \cite{edwards} that is termed \emph{unitary}
(or quasi-continuous), since the control terms are divided by the
norm of the states. In \cite{vidal1} conditions on the (unitary)
MGSTA gains are given to ensure the global finite-time convergence
despite of an unknown but \textbf{constant} and symmetric input matrix
and Lipschitz perturbations. In \cite{vidal2} a (unitary) MGSTA
with time and state-dependent gains is proposed for systems with known
input matrix and state-dependent perturbations. However, they have
not taken into account the main problem appearing in this case: in
the derivative of the state-varying perturbation, the control signal
appears once more (see the motivation example in \cite{isma} and Section \ref{sec:ME}), causing
an algebraic loop that requires special treatment. 

In synthesis, the existing MGSTA designs cannot ensure the convergence
of the trajectories of the system in three very important situations: 
\begin{itemize}
\item Systems with a state- and time-dependent uncertain input matrix, \textit{e.g.,}
mechanical systems with parametric uncertainties in the inertia matrix. 
\item Systems with state- and time-dependent perturbations, \textit{e.g.,}
the tracking problem of any uncertain mechanical system. 
\item Systems with state- and time-dependent actuator faults. 
\end{itemize}

In this paper a unitary MGSTA for the control of a system with state- and time-dependent uncertain control matrix and perturbations is
proposed. It is shown that under reasonable assumptions for the uncertainties,
it is always possible to find a set of constant gains to ensure global
and robust finite-time stability of the closed-loop system. Furthermore,
a procedure to choose a set of gains is proposed. Moreover,
the structure presented in \cite{edwards,vidal1} and \cite{vidal2} is generalized
by allowing not only discontinuous terms in the algorithm, adding
an extra degree of freedom in the design.

The tracking control problem for an omnidirectional mobile robot is
considered as a motivation example. In this example it is shown that
the conditions given by \cite{edwards,vidal1} and \cite{vidal2} cannot be satisfied.
On the other hand, the proposed methodology ensures the global finite-time
stability of the closed-loop system for a larger class of dynamical
systems. The simulation results illustrate the effectiveness of the
proposed methodology.

This paper is structured as follows: Section \ref{sec:ME} presents
a motivational example, while Section \ref{sec:PS} states the problem
to be solved and working assumptions. In Section \ref{sec:MR} the
main result of the paper is announced. Section \ref{sec:SR} shows
a simulation study illustrating the results and Section \ref{sec:CR}
furnishes some concluding remarks. The stability proof of the main
result is provided in the Appendix.

\textbf{Notation:} For a matrix $M\in\mathbb{R}^{n\times n}$, the
symmetric component is $\mbox{{\rm Sym}\ensuremath{\left\{ M\right\} } }=\frac{1}{2}\left(M+M^{T}\right)$,
while $\mbox{{\rm Gram}\ensuremath{\left\{ M\right\} } }=M^{T}M$
is the Gramian matrix. Recall that $\left\Vert M\right\Vert _{2}^{2}=\sigma_{\max}^{2}\left\{ M\right\} =\lambda_{\max}\left\{ {\rm Gram}\left\{ M\right\} \right\} $,
where $\sigma_{\max}\left\{ M\right\} $ represents the largest singular
value of matrix $M$ and $\mathbb{R}_{+}=\left\{ x\in\mathbb{R}\mid x\geq0\right\}$.

\section{Motivation Example}
\label{sec:ME} 
Consider the dynamics of a four-wheeled omnidirectional
mobile robot (for more details, see \cite{b:Campion_IEEE_TRA1996}
and \cite{kelly}): 
\begin{equation}\small{\label{eq:rob}
\ddot{q}=M^{-1}\left[\frac{k_ar_e}{r_a}R^T(\theta)E\nu - C(\dot{q})\dot{q}
- f_r(\dot{q}) + w(t,q,\dot{q})\right],}
\end{equation}\normalsize
where $q=[x,y,\theta]^{T}\in\mathbb{R}^{3}$ is a vector containing
the configuration variables in the task space, $\nu\in\mathbb{R}^{4}$
is a vector with the motor armature voltages while $w:\mathbb{R}\times\mathbb{R}^{3}\times\mathbb{R}^{3}\to\mathbb{R}^{3}$
represents some external disturbances. The matrices $M\in\mathbb{R}^{3\times3}$,
$R(\theta)\in\mathbb{R}^{3\times3}$, $E\in\mathbb{R}^{3\times4}$,
and $C(\dot{q})\in\mathbb{R}^{3\times3}$ are the inertia matrix,
the rotation matrix for a planar motion, the transpose of the Jacobian
matrix and the Coriolis/centrifugal matrix, respectively; while $f_{r}:\mathbb{R}^{3}\to\mathbb{R}^{3}$
is a friction force vector containing viscous and dry friction.

The structure of the system matrices is given by: 
\begin{gather*}
M=\textrm{diag}\{M_{1},M_{2},M_{3}\}+(J_{2}+J_{m}r_{e}^{2})EE^{T},\\
R(\theta)=\left[\begin{array}{ccc}
\cos\theta & \sin\theta & 0\\
-\sin\theta & \cos\theta & 0\\
0 & 0 & 1
\end{array}\right],\\
E=\frac{1}{r}\left[\begin{array}{cccc}
1 & 1 & 1 & 1\\
1 & -1 & 1 & -1\\
L & -L & -L & L
\end{array}\right],\\
C(\dot{q})=\frac{4}{r^{2}}(J_{2}+J_{m}r_{e}^{2})\tanh(\dot{\theta})B,\ B=\left[\begin{array}{ccc}
0 & 1 & 0\\
-1 & 0 & 0\\
0 & 0 & 0
\end{array}\right],\\
f_{r}(\dot{q})=f_{v}(\dot{q})+f_{d}(\dot{q}),\\
f_{v}(\dot{q})=\left[\begin{array}{c}
f_{vx}\dot{x}\\
f_{vy}\dot{y}\\
f_{v\theta}\dot{\theta}
\end{array}\right],\ f_{d}(\dot{q})=\left[\begin{array}{c}
f_{dx}\tanh(\dot{x})\\
f_{dy}\tanh(\dot{y})\\
f_{d\theta}\tanh(\dot{\theta})
\end{array}\right],
\end{gather*}
where $M_{1}=M_{2}=m_{1}+4m_{2}$, $M_{3}=4m_{2}(l_{1}^{2}+l_{2}^{2})+J_{1}+4J_{3}$,
with $m_{1}$ being the mass of the body, $m_{2}$ the mass of each
wheel, $J_{1}$ the inertia of the body, $J_{2}$ and $J_{3}$ the
inertia of the wheels over and perpendicular to the motor shaft, respectively;
$l_{1}$ and $l_{2}$ the length from the robot center to the robot
front and robot side, respectively; $r$ the wheels radius, $L$ the
distance from the robot center to the wheels, $J_{m}$ the inertia
of the motor shaft, $k_{a}$ the torque constant, $r_{a}$ the armature
resistance and $r_{e}$ the gear ratio.

Let us consider the tracking problem; thus, the tracking error is
defined as $\tilde{q}=q-q_{d}(t),$ where $q_{d}(t)$ is a reference
signal to be tracked. Then, in order to control system (\ref{eq:rob})
via the MGSTA a sliding variable is designed as 
\[
s=\Theta\tilde{q}+\dot{\tilde{q}},
\]
where $\Theta\in\mathbb{R}^{3\times3}$ is a positive definite matrix
to be designed. Suppose that only nominal values for $M$, $k_{a}$,
$r_{a}$ and $r_{e}$ are available. Thus, the dynamics for $s$ can
be written as 
\begin{multline*}
\dot{s}=\Theta\dot{\tilde{q}}-\ddot{q}_{d}+\bar{M}(q)u+M^{-1}[-C(\dot{q})\dot{q}\\
-f_{r}(\dot{q})+w(t,q,\dot{q})],
\end{multline*}
where $u=E\nu$ and $\bar{M}(q)=k_{a}r_{e}M^{-1}R^{T}(\theta)/r_{a}$.
Taking into account that the friction forces $f_{v}$ and $f_{c}$
are unknown; then, they represent some additional uncertainties/perturbations
on the system. Additionally, the matrix $\bar{M}(q)$ is uncertain
in the sense that there exist some matrices $M_{0}(q)$ and $\Delta_{M}(q)$
such that $\bar{M}(q)=(I+\Delta_{M}(q))M_{0}(q)$, where $M_{0}(q)$
and $\Delta_{M}(q)$ represent a nominal known and an unknown part
of the matrix $\bar{M}(q)$, respectively. Note that, when $\Delta_{M}(q)=0$,
the nominal case is recovered, and for $M_{0}(q)=I$ and $\Delta_{M}(q)=\bar{M}(q)-I$,
the case with no available knowledge is considered. This representation
becomes meaningful in contexts where the inertial parameters may vary,
\textit{e.g.,} when the robot changes mass due to the task, or in
tasks where actuator faults may be present, to list only a few examples.

Thus, if the control law $u=u_{st}$ is considered, with $u_{st}$
representing any MGSTA; then, the closed-loop system dynamics can
be rewritten as 
\begin{multline}
\dot{s}=\Theta\dot{\tilde{q}}-\ddot{q}_{d}+(I+\Delta_{M}(q))M_{0}(q)u_{st}\\
+M^{-1}[-C(\dot{q})\dot{q}-f_{r}(\dot{q})+w(t,q,\dot{q})].\label{eq:S_dynamics}
\end{multline}

Note that the term $\Delta_{M}(q)M_{0}(q)u_{st}$ cannot be canceled;
thus, such a term needs to be considered as a part of the perturbation.
This issue presents a challenge since the perturbation would depend
directly on the control law and the gain selection cannot be made
without assuming the boundedness of the trajectories of the closed-loop
system \textit{a priori}.

Furthermore, under the assumptions made in \cite{moreno} and \cite{vidal2},
which represent the most general conditions presented on the literature,
the only way to deal with the dry friction model is to assume that
it has a bounded time derivative, \textit{i.e.,} that 
\[
\dt f_{d}(\dot{q})=\left[\begin{array}{c}
f_{dx}(1-\tanh^{2}(\dot{x}))\ddot{x}\\
f_{dy}(1-\tanh^{2}(\dot{y}))\ddot{y}\\
f_{d\theta}(1-\tanh^{2}(\dot{\theta}))\ddot{\theta}
\end{array}\right],
\]
is bounded. However, note that this term depends linearly on the control
signal. Therefore, to design the controller gains, the bounds of the
accelerations need to be known \textit{a priori}. It is important
to remark that the works presented in \cite{edwards,vidal1,vidal2,moreno}
do not consider this algebraic loop.

\section{Problem Statement}
\label{sec:PS} 
Consider the following dynamical system 
\begin{equation}
\dot{x}=f\left(t,x\right)+G\left(t,x\right)u,\label{eq:sys}
\end{equation}
where $x\in\mathbb{R}^{n}$ represents the state vector, $u\in\mathbb{R}^{n}$
is the control input, $f\left(t,x\right)\in\mathbb{R}^{n}$ is an
uncertain vector field containing uncertainties and/or perturbations,
and $G\left(t,x\right)\in\mathbb{R}^{n\times n}$ is the uncertain
control (or input) matrix. 

The uncertain input matrix $G\left(t,x\right)$ will be represented
as 
\begin{equation}
G\left(t,x\right)=\left(I+\Delta_{G}\left(t,x\right)\right)G_{0}\left(t,x\right),\label{eq:GMatrix}
\end{equation}
where $G_{0}:\mathbb{R}_{+}\times\mathbb{R}^{n}\rightarrow\mathbb{R}^{n\times n}$
corresponds to the nominal part, which is assumed to be invertible,
and $\Delta_{G}:\mathbb{R}_{+}\times\mathbb{R}^{n}\rightarrow\mathbb{R}^{n\times n}$
is the uncertain term. 

In order to control (\ref{eq:sys}), following \cite{edwards} and \cite{moreno},
an MGSTA in the form 
\begin{subequations} \label{eq:GSTA} 
\begin{align}
u & =-k_{1}\phi_{1}\left(x\right)+bG_{0}^{-1}\left(t,x\right)v,\\
\dot{v} & =-k_{2}\phi_{2}\left(x\right),
\end{align}
\end{subequations}
is proposed. The non-linear functions $\phi_{i}:\mathbb{R}^{n}\rightarrow\mathbb{R}^{n}$ are monotonically increasing, $\phi_{1}$ is continuous everywhere and $\phi_{2}$
is continuous everywhere except possibly at $x=0$. We consider them
as given by 
\begin{align}
\phi_{1}(x) & =(\alpha\left\Vert x\right\Vert ^{-p}+\beta)x,\label{eq:ST1}\\
\phi_{2}(x) & =J(x)\phi_{1}(x)=c(x)\phi_{1}(x),\label{eq:ST2}
\end{align}
where $J(x)$ is the Jacobian matrix of $\phi_{1}(x)$
and $c(x)$ is a scalar function, which are defined
in $\mathbb{R}^{n}\setminus\left\{ 0\right\}$, \textit{i.e.,}
\begin{align}
J(x) & =\frac{\partial\phi_{1}(x)}{\partial x}=(\alpha\left\Vert x\right\Vert ^{-p}+\beta)I-\alpha p\frac{\left\Vert x\right\Vert ^{-p}}{\left\Vert x\right\Vert ^{2}}xx^{T},\label{eq:J}\\
c(x) & =\alpha(1-p)\left\Vert x\right\Vert ^{-p}+\beta.\label{eq:c}
\end{align}

The (positive) constant scalar gains $k_{1},\,k_{2}\in\mathbb{R}_{+}$
have to be designed, while the control coefficient gain $b>0$, the
internal positive gains $\alpha>0,\,\beta>0$ and the power $p\in\left(0,\,\frac{1}{2}\right]$
can be freely chosen by the designer. Note that the relation $\phi_{2}(x)=J(x)\phi_{1}(x)$
in (\ref{eq:ST2}) is an extension to the multivariable case of the
scalar relation $\phi_{2}(x)=\phi'_{1}(x)\phi_{1}(x)$
(see \cite{moreno_osorio} and \cite{moreno} for further details). 
\begin{rem}
\label{rem:Jispd}
\textit{The properties of matrix $\mathcal{J}(x)\triangleq
\frac{1}{c(x)}J(x)$ are important. It is easily seen that it is symmetric since $J(x)=J^{T}(x)$. It is continuous in $\mathbb{R}^{n}\setminus\left\{ 0\right\}$,
and in contrast to $J\left(x\right)$ it is bounded in $\mathbb{R}^{n}$,
although its limit for $x=0$ does not exist. Moreover, it has one
eigenvalue at $\lambda=1$, with eigenvector $x$. It has also $\left(n-1\right)$
repeated eigenvalues with value $1\leq\lambda=\frac{\alpha+\beta\left\Vert x\right\Vert ^{p}}{\alpha\left(1-p\right)+\beta\left\Vert x\right\Vert ^{p}}<\frac{1}{1-p}$, with corresponding linearly independent eigenvectors in each of the
$\left(n-1\right)$ different directions orthogonal to the vector
$x$. Since $\mathcal{J}$ is symmetric and all eigenvalues are positive and upper and lower bounded,
it follows that $\mathcal{J}\left(x\right)$ is positive definite
for all $x\in\mathbb{R}^{n}$,} i.e.,
\[
I\leq\mathcal{J}\left(x\right)\triangleq\frac{1}{c\left(x\right)}J\left(x\right)\leq\frac{\alpha+\beta\left\Vert x\right\Vert ^{p}}{\alpha\left(1-p\right)+\beta\left\Vert x\right\Vert ^{p}}I\,.
\]
\end{rem}

It is well-known \cite{edwards,vidal1,vidal2,moreno} that in the
absence of input matrix uncertainty, \textit{i.e.,} $\Delta_{G}\equiv0$, and
for some bounds on the perturbation terms in $f\left(t,x\right)$
the origin of the closed-loop system (\ref{eq:sys})-(\ref{eq:GSTA})
is finite-time stable by an appropriate selection of the gains $k_{1},\,k_{2}$.
The objective of this paper is to find conditions on the uncertainty
of the input matrix and on $f\left(t,x\right)$ such that the robust
finite-time stability of the closed-loop system is assured. 

For this, the uncertain term $f(t,x)$ is assumed to be
decomposed as 
\begin{equation}\small{
f\left(t,x\right)=f_{1}(t,x)+f_{2}(t,x)=\Delta_{1}(t,x)\phi_{1}(x)+f_{2}(t,x),\label{eq:f}}
\end{equation}\normalsize
where $f_{2}\left(t,x\right)$ is such that 
\begin{equation}
\label{eq:df2}
\frac{d}{dt}\left[\left(I+\Delta_{G}\right)^{-1}f_{2}\right]=\Delta_{2}\left(t,x\right)\phi_{2}\left(x\right)+\Delta_{3}\left(t,x\right)\dot{x}.
\end{equation}

Moreover, introduce the following assumptions on the input matrix
and the perturbations. 
\begin{assumption}\label{A:fs} For the dynamical system (\ref{eq:sys})
and controller (\ref{eq:ST1})-(\ref{eq:ST2}), the following conditions
hold for all $t\geq0$ and $x\in\mathbb{R}^{n}$: 
\begin{enumerate} 
\item There exist some positive constants $0<g_{m}^{\Delta}\leq g_{M}^{\Delta}$
such that 
\begin{equation}
g_{m}^{\Delta}I\leq G\left(t,x\right)+G^{T}\left(t,x\right)\leq g_{M}^{\Delta}I.\label{eq:Gbounds}
\end{equation}
\item Matrices $\Delta_{i}$, $i=1,\,2,\,3$, in \eqref{eq:f} and \eqref{eq:df2}, are bounded, \textit{i.e.,} there
exist $\delta_{i}\geq0$ such that
\[
\left\Vert \Delta_{i}\left(t,x\right)\right\Vert \leq\delta_{i},\,i=1,\,2,\,3.
\]
\item There exist some positive constants $0<\gamma_{1}^{\Delta}\leq\gamma_{2}^{\Delta}$
such that 
\begin{equation}
0<\gamma_{1}^{\Delta}I\leq2{\rm Sym}\left\{ \mathcal{J}\left(x\right)\left(I+\Delta_{G}\right)\right\} \leq\gamma_{2}^{\Delta}I.\label{eq:BoundDeltaG}
\end{equation}
\end{enumerate} 
\end{assumption}

For future reference, note that as a consequence of the previous assumptions
there also exist non-negative constants $\gamma_{3},\,\gamma_{4},\,\gamma_{5}$
such that
\begin{subequations}
\label{eq:MainBounds2}
\begin{align}
{\rm Gram}\left\{ \mathcal{J}G\right\}  & \leq\gamma_{3}I, \\
2{\rm Sym}\left\{ \Delta_{G}\mathcal{J}G\right\}  & \leq\gamma_{4}I,\\
{\rm Gram}\left\{ \Delta_{G}^{T}\right\}  & \leq\gamma_{5}I.
\end{align}
\end{subequations}

Some comments on Assumption \ref{A:fs} are in order. 
\begin{itemize}
\item Conditions 1) and 3) refer to the size of $\Delta_{G}$. They require
$G_{0}$ to have a positive definite symmetric part, and $\Delta_{G}$
to be sufficiently small not to destroy this property either for $G$
nor for $\mathcal{J}$. These conditions are clearly weaker than those
considered in the previous papers \cite{edwards,vidal1,vidal2} and \cite{moreno}.
\item Condition 2) implies that the term $f_{1}(t,x)$ contains vanishing
perturbations, while $f_{2}(t,x)$ can have non-vanishing disturbances.
Moreover, condition 2) considers that the derivative of the perturbation
depends on $\dot{x}$ and thus of the control signal. This term needs
to be considered to properly manage the algebraic loop mentioned in
the motivational example and in \cite{isma}. 
\item Note that the depending on the class of perturbations, an appropriate
value of $p$ needs to be selected. If \textit{e.g.,} the derivative of $f_{2}$
has terms not vanishing at zero, then $p=\frac{1}{2}$ will be suitable,
as is considered in \cite{edwards,vidal1} and \cite{vidal2}. Other values of
$p$ can be considered though.
\end{itemize}

\section{Main Result}
\label{sec:MR} 
In this section sufficient conditions for the stability
of the closed-loop system (\ref{eq:sys})-(\ref{eq:GSTA}) are derived.
Moreover, a procedure to find gains $k_{1},\,k_{2}$ assuring the
stability is provided. 

For convenience, we drop the arguments of the functions, when this
does not lead to confusion. Defining the variable $z=v+b^{-1}\left(I+\Delta_{G}\right)^{-1}f_{2}$
the closed-loop system (\ref{eq:sys})-(\ref{eq:GSTA}) can be written
as 
\begin{align}
\dot{x} & =f_{1}+\left(I+\Delta_{G}\right)\left(-G_{0}k_{1}\phi_{1}+bz\right),\label{eq:m1}\\
\dot{z} & =-k_{2}\phi_{2}+\Delta_{2}\phi_{2}+\Delta_{3}\dot{x}\,.\label{eq:m2}
\end{align}

The following theorem is the main result of the paper and ensures
the global and robust finite-time stability of the closed-loop system
(\ref{eq:m1})-(\ref{eq:m2}).
\begin{theorem}
\label{T:gains}
Select $\alpha>0$, $\beta>0$, $b>0$ and $p\in(0,\frac{1}{2}]$
and let Assumption \ref{A:fs} be satisfied. Assume further that the
inequality 
\begin{equation}
\gamma_{1}^{\Delta}g_{m}^{\Delta}>\gamma_{4}+2\sqrt{\gamma_{3}\gamma_{5}},\label{eq:FeasibCond}
\end{equation}
is satisfied. Then, for arbitrary values of $\delta_{1}$, $\delta_{2}$,
$\delta_{3}$, there exist positive constant controller gains $k_{1},\,k_{2}$
such that $[x^T,z^T]^{T}=0$ is globally robust finite-time stable. 
\end{theorem}

The proof of the Theorem is given in the Appendix, using a (non-smooth)
Lyapunov function. Note that the key condition (\ref{eq:FeasibCond})
depends only on the size of the uncertainty of the control matrix
$\Delta_{G}$. If $\Delta_{G}\equiv0$ then (\ref{eq:FeasibCond})
is satisfied, since $\gamma_{4}=\gamma_{5}=0$. So we come to the
rather surprising conclusion that if the closed-loop can be stabilized
for $f\equiv0$ in presence of the uncertainty $\Delta_{G}$, it
can be stabilized for arbitrary large $f$ satisfying Assumption \ref{A:fs}.

Theorem 1 generalizes the results presented in \cite{edwards,vidal1,vidal2} and \cite{moreno}
by considering the presence of an uncertain time- and state-dependent
input matrix. Moreover, the algebraic loop caused by the presence
of $\dot{x}$ in the perturbation, and discused in detail in \cite{isma},
have been appropriately taken into account.

Note that after a finite time $z\equiv0$, so that $v\left(t\right)\equiv-b^{-1}\left(I+\Delta_{G}\left(t,x\right)\right)^{-1}f_{2}\left(t,x\right)$.
This shows that the term $v$ of the MGSTA (\ref{eq:GSTA}) estimates
exactly the (negative) value of the perturbation, and can therefore
compensate for it.

To complete the paper, a procedure to select stabilizing gains $k_{1},\,k_{2}$
will be given. Assumption \ref{A:fs} assures the existence of constants
$\mu_{1},...,\mu_{4}$ and $\theta_{1},...,\theta_{12}$ such that
the following bounds are satisfied: 
\begin{subequations} \label{eq:DefMus} 
\begin{align}
2{\rm Sym}\left\{ \Delta_{1}\right\}  & \leq\mu_{1}I,\\
2{\rm Sym}\left\{ AG\right\}  & \leq\mu_{2}I,\\
2{\rm Sym}\left\{ A\Delta_{1}+\Delta_{2}\right\}  & \leq\mu_{3}I,\\
2{\rm Sym}\left\{ A\left(I+\Delta_{G}\right)\right\}  & \leq\mu_{4}I,
\end{align}
\end{subequations} 
with $A=\frac{\Delta_{3}}{c}$, and
\begin{subequations}
\label{eq:DefThetas}
\small{
\begin{align}
\theta_{1}I&\geq2{\rm Sym}\{ G^{T}\mathcal{J}AG\},   \\
\theta_{2}I&\geq{\rm Gram}\{ AG\},\\
\theta_{3}I&\geq2{\rm Sym}\{[A(I+\Delta_{G})b+\Delta_{1}^{T}\mathcal{J}]\mathcal{J}G\},\\
\theta_{4}I&\geq2{\rm Sym}\{(\Delta_{2}^{T}+\Delta_{1}^{T}A^{T})\mathcal{J}G\},\\
\theta_{5}I&\geq2{\rm Sym}\{[A(I+\Delta_{G})b+\Delta_{1}^{T}\mathcal{J}]AG\},\\
\theta_{6}I&\geq2{\rm Sym}\{\Delta_{G}AG\},\\
\theta_{7}I&\geq2{\rm Sym}\{(\Delta_{2}^{T}+\Delta_{1}^{T}A^{T})AG\},\\
\theta_{8}I&\geq2{\rm Sym}\{\Delta_{G}[(I+\Delta_{G}^{T})A^{T}b+\mathcal{J}\Delta_{1}]\},\\
\theta_{9}I&\geq2{\rm Sym}\{\Delta_{G}(\Delta_{2}+A\Delta_{1})\},\\
\theta_{10}I&\geq2{\rm Sym}\{(\Delta_{2}^{T}+\Delta_{1}^{T}A^{T})[(I+\Delta_{G}^{T})A^{T}b+\mathcal{J}\Delta_{1}]\},\\
\theta_{11}I&\geq{\rm Gram}\{\Delta_{2}+A\Delta_{1}\},\\
\theta_{12}I&\geq{\rm Gram}\{(I+\Delta_{G}^{T})A^{T}b+\mathcal{J}\Delta_{1}\}. 
\end{align}}
\end{subequations}\normalsize
\indent For positive $p_{1}>0$ and $p_{2}>0$, define the functions
{\small{}{ 
\begin{align*}
\Xi_{1}\left(p_{1},p_{2}\right) & =\theta_{8}+\theta_{9}p_{2}+\theta_{10}\frac{p_{2}}{bp_{1}}+\theta_{11}\frac{p_{2}^{2}}{bp_{1}}+\theta_{12}\frac{1}{bp_{1}},\\
\Xi_{2}\left(p_{2}\right) & =\frac{1}{b}\left(\theta_{3}+\left(\theta_{4}+\theta_{5}\right)p_{2}+\theta_{7}p_{2}^{2}\right),\\
\Xi_{3}\left(p_{2}\right) & =\frac{1}{b}\left(\gamma_{3}+\theta_{1}p_{2}+\theta_{2}p_{2}^{2}\right),\\
\Gamma_{0}\left(p_{1},p_{2}\right) & =\left(\Xi_{2}+\left(\gamma_{1}^{\Delta}-p_{2}\mu_{4}\right)\mu_{2}\right)\frac{1}{p_{1}}+\left(\theta_{6}+\mu_{4}g_{m}^{\Delta}\right)p_{2},\\
\Gamma_{1}\left(p_{1},p_{2}\right) & =\left(\Gamma_{0}+\gamma_{4}-\gamma_{1}^{\Delta}g_{m}^{\Delta}\right)^{2}-4b\Xi_{3}\gamma_{5},\\
\Gamma_{2}\left(p_{1},p_{2}\right) & =4\Xi_{3}\left\{ \left(\frac{2b}{p_{2}}+\mu_{1}+\frac{\mu_{3}}{p_{1}}\right)\left(\gamma_{1}^{\Delta}-p_{2}\mu_{4}\right)+\Xi_{1}\right\} ,
\end{align*}
}}and also
\begin{subequations} \label{eq:DefAlphas} 
\begin{align}
\alpha_{2} & =\frac{1}{\tilde{\gamma}_{1}^{\Delta}}\Xi_{3}\left(p_{2}\right),\\
\alpha_{1} & =\frac{1}{\tilde{\gamma}_{1}^{\Delta}}\left(\Xi_{2}+\gamma_{4}p_{1}+\theta_{6}p_{1}p_{2}\right)+\mu_{2}-g_{m}^{\Delta}p_{1},\\
\alpha_{0} & =2b\frac{p_{1}}{p_{2}}+\mu_{1}p_{1}+\mu_{3}+\frac{1}{b\tilde{\gamma}_{1}^{\Delta}}\left(\gamma_{5}bp_{1}+\Xi_{1}\right)bp_{1},
\end{align}
\end{subequations}
with $\tilde{\gamma}_{1}^{\Delta}=\gamma_{1}^{\Delta}-p_{2}\mu_{4}$.
The design of the gains follows the following algorithm, which is
derived from the proof of Theorem \ref{T:gains} in Appendix. 

\begin{alg}\label{A:gs}In order to ensure a proper selection of
the gains $k_{1}$ and $k_{2}$ in (\ref{eq:GSTA}), consider the
following steps: 
\begin{enumerate} 
\item Select $\alpha>0$, $\beta>0$, $b>0$ and $p\in(0,\frac{1}{2}]$.
The parameter $p$ should be selected depending on the size of the
perturbations. 
\item Find a value of $p_{2}=p_{2}^{*}$ and a value of $p_{1}=p_{1}^{*}$
such that 
\begin{equation}
\gamma_{1}^{\Delta}>p_{2}\mu_{4},\label{eq:Cond1p2}
\end{equation}
and 
\begin{equation}
\gamma_{1}^{\Delta}g_{m}^{\Delta}-\gamma_{4}-\Gamma_{0}\left(p_{1},p_{2}\right)>0,\label{eq:IneqPModb}
\end{equation}
are satisfied for $p_{2}=p_{2}^{*}$ and $p_{1}\geq p_{1}^{*}$. These
values always exist. 
\item Find a value $p_{1}^{\#}$ of $p_{1}$ such that $p_{1}^{\#}\geq p_{1}^{*}$
and that the following inequality 
\begin{equation}
\Gamma_{1}\left(p_{1},p_{2}\right)p_{1}>\Gamma_{2}\left(p_{1},p_{2}\right),\label{eq:IneqPMod}
\end{equation}
is satisfied for every $p_{1}\geq p_{1}^{\#}$. This is also always
feasible. 
\item Fix $p_{2}=p_{2}^{*}$, $p_{1}\geq p_{1}^{\#}$ and $p_{1}p_{2}>1$,
choose 
\begin{equation}
k_{2}=b\frac{p_{1}}{p_{2}},\label{eq:Selectionk2}
\end{equation}
and 
\begin{equation}
k_{1}\in\left(\frac{-\alpha_{1}-\bar{\alpha}}{2\alpha_{2}},\frac{-\alpha_{1}+\bar{\alpha}}{2\alpha_{2}}\right),\label{eq:Selectionk1}
\end{equation}
with $\bar{\alpha}=\sqrt{\alpha_{1}^{2}-4\alpha_{2}\alpha_{0}}$. 
\end{enumerate}
\end{alg}

\section{Examples}
\label{sec:SR} 
\subsection{An Academic Example}
Consider a simple situation with $n=2$ and in (\ref{eq:GMatrix})
\begin{align*}
G_{0} & =I_{2},\,\Delta_{G}=\left[\begin{array}{cc}
0 & g\left(t\right)\\
g\left(t\right) & 0
\end{array}\right]\,,
\end{align*}
with $g\left(t\right)$ an arbitrary time-varying signal with $\left|g\left(t\right)\right|\leq\bar{g}$,
that represents the coupling between the two control channels. We
calculate the value of $\bar{g}$ such that the key condition (\ref{eq:FeasibCond})
is satisfied, \textit{i.e.,} there exist stabilizing gains. Unfortunately, even
in this simple case, it seems to be impossible to obtain an analytical
expression for the condition, due to the discontinuous nature of $\mathcal{J}$.
However, a good approximation can be computed numerically. For $\alpha=\beta=1$,
$p=1/2$, $g_{m}^{\Delta}=2\left(1-\bar{g}\right)$, $\gamma_{5}=\bar{g}^{2}$,
and for $\bar{g}=0.23$ condition (\ref{eq:FeasibCond}) is still
fulfilled. This is possibly a conservative estimation of the true
value, and so are also the gains calculated by the Algorithm \ref{A:gs}.

\subsection{The Motivation Example}
For the four-wheeled omnidirectional mobile robot closed-loop dynamics
(\ref{eq:S_dynamics}), the representation in (\ref{eq:sys}) is given
as follows: 
\begin{align*}
f(t,s) & =\Theta\dot{\tilde{q}}-\ddot{q}_{d}+M^{-1}[-C(\dot{q})\dot{q}-f_{r}(\dot{q})+w(t,q,\dot{q})],\\
G(t,s) & =(I+\Delta_{M}(q))M_{0}(q).
\end{align*}

Recalling that $s=\Theta\tilde{q}+\dot{\tilde{q}}$, that the friction
forces $f_{v}$ and $f_{d}$ are unknown, and that the matrix $\bar{M}(q)$
is uncertain, it follows that 
\begin{align*}
f_{1}(t,s) & =M^{-1}[M\Theta-C(\dot{q})-\bar{f}_{v}]s+w_{1}(t,s),\\
f_{2}(t,s) & =M^{-1}[C(\dot{q})\Theta+\bar{f}_{v}-M\Theta^{2}]\tilde{q}-\ddot{q}_{d}\\
 & -M^{-1}[C(\dot{q})+\bar{f}_{v}]\dot{q}_{d}-M^{-1}f_{d}(\dot{q})+w_{2}(t,s),\\
\Delta_{M}(q) & =[\bar{M}(q)-M_{0}(q)]M_{0}^{-1}(q),\\
M_{0}(q) & =\frac{k_{an}r_{en}}{r_{an}}M_{n}^{-1}R^{T}(\theta),\ \bar{M}(q)=\frac{k_{a}r_{e}}{r_{a}}M^{-1}R^{T}(\theta),
\end{align*}
where $\bar{f}_{v}=\textrm{diag}\{f_{vx},f_{vy},f_{v\theta}\}$ is
a viscous friction matrix; $M_{n}$, $k_{an}$, $r_{an}$ and $r_{en}$
are nominal values for $M$, $k_{a}$, $r_{a}$ and $r_{e}$, respectively;
and the external disturbances $w_{1}$ and $w_{2}$ are such that
$w(t,s)=w_{1}(t,s)+w_{2}(t,s)$.

Therefore, one can verify that, for all $\theta\in(-\pi/2,\pi/2)$,
condition 1) of Assumption \ref{A:fs} holds for some $g_{m}^{\Delta},g_{M}^{\Delta}>0$. Additionally, $f_{1}$ satisfies 
\begin{align*}
f_{1}(t,s) & =\Delta_{1}(t,s)(\alpha||s||^{-p}s+\beta s),\\
\Delta_{1}(t,s) & =M^{-1}[M\Theta-C(\dot{q})-\bar{f}_{v}],\\
w_{1}(t,s) & =\alpha\Delta_{1}(t,s)||s||^{-p}s,
\end{align*}
with $\alpha>0$ and $\beta=1$. Then, due to the properties of the
omnidirectional mobile robot, one could assume that $||f_{1}(t,s)||\leq\delta_{1}||\phi_{1}(s)||$,
for some $\delta_{1}>0$.

On the other hand, assuming that $\dot{q}_{d}(t)=cte.$, and $M_{n}=\lambda_{M}I$
with some $\lambda_{M}>0$; after some calculations, $\bar{f}_{2}(t,s)=(I+\Delta_{M}(q))^{-1}f_{2}(t,s)$
satisfies{\small{}{ 
\begin{align*}
\frac{d}{dt}\bar{f}_{2}(t,s) & =\Delta_{2}(t,s)[\alpha^{2}(1-p)||s||^{-2p}s\\
 & \qquad+\alpha\beta(2-p)||s||^{-p}s+\beta^{2}s]+\Delta_{3}(t,s)\dot{s},\\
\Delta_{2}(t,s) & =(I+\Delta_{M}(q))^{-1}M^{-1}[C(\dot{q})\Theta+\bar{f}_{v}-M\Theta^{2}],\\
\Delta_{3}(t,s) & =(I+\Delta_{M}(q))^{-1}M^{-1}\left[\frac{4}{r^{2}}(J_{2}+J_{m}r_{e}^{2})B\times\right.\\
 & (1-\tanh^{2}(\dot{\theta}))(\Theta\tilde{q}-\dot{q}_{d})\bar{\mathbf{1}}-\bar{f}_{d}(I-\tanh^{2}(\dot{q}))\Big],\\
\frac{d}{dt}w_{2}(t,s) & =\alpha\Delta_{2}(t,s)[\alpha(1-p)||s||^{-2p}+\beta(2-p)||s||^{-p}]s,
\end{align*}
}}with $\bar{\mathbf{1}}=[0,0,1]$, $\bar{f}_{d}=\textrm{diag}\{f_{dx},f_{dy},f_{d\theta}\}$
is a dry friction matrix; $\alpha>0$ and $\beta=1$. Thus, due to
the properties of the omnidirectional mobile robot, one could assume
that $||d\bar{f}_{2}(t,s)/dt||\leq\delta_{2}||\phi_{2}(s)||+\delta_{3}||\dot{s}||$,
for some $\delta_{2},\delta_{3}>0$. Hence, condition 2) of Assumption
\ref{A:fs} is satisfied.

Finally, due to the fact that $\mathcal{J}(s)$ is positive definite and $\theta\in(-\pi/2,\pi/2)$,
condition 3) of Assumption \ref{A:fs} holds for some $\gamma_{1}^{\Delta},\gamma_{2}^{\Delta}>0$.

In order to illustrate the performance of the MGSTA (\ref{eq:GSTA}), some simulations
are carried out in Matlab with Euler explicit discretization method
and sampling time equal to $1\times10^{-3}$. The parameters of the
mobile robot are given in Table \ref{table:parameters}. The units
for $f_{vx}$, $f_{vy}$ and $f_{v\theta}$ are $\mathrm{[kg/s]}$,
$\mathrm{[kg/s]}$ and $\mathrm{[kgm^{2}/s]}$, respectively; while
for $f_{dx}$, $f_{dy}$ and $f_{d\theta}$ are $\mathrm{[kgm/s^{2}]}$,
$\mathrm{[kgm/s^{2}]}$ and $\mathrm{[kgrad/s^{2}]}$, respectively.
The nominal values for $M$, $k_{a}$, $r_{a}$ and $r_{e}$ are taken
as $M_{n}= 7.7341I$, $k_{an}=0.0130$, $r_{an}=1.9$ and $r_{en}=57$.

\begin{table}
\centering \caption{System Parameters}
{\scriptsize{}{ }%
\begin{tabular}{|c|c|c|c|}
\hline 
\textbf{\scriptsize{}Parameter} & \textbf{\scriptsize{}Value} & \textbf{\scriptsize{}Parameter} & \textbf{\scriptsize{}Value}\tabularnewline
\hline 
\hline 
{\scriptsize{}$m_{1}$} & {\scriptsize{}$2.8[\mathrm{kg}]$} & {\scriptsize{}$r$} & {\scriptsize{}$0.042[\mathrm{m}]$}\tabularnewline
\hline 
{\scriptsize{}$m_{2}$} & {\scriptsize{}$0.38[\mathrm{kg}]$} & {\scriptsize{}$J_{m}$} & {\scriptsize{}$5.7e^{-7}[\mathrm{kgm^{2}}]$}\tabularnewline
\hline 
{\scriptsize{}$J_{1}$} & {\scriptsize{}$0.0608[\mathrm{kgm^{2}}]$} & {\scriptsize{}$k_{a}$} & {\scriptsize{}$0.013[\mathrm{Nm/A}]$}\tabularnewline
\hline 
{\scriptsize{}$J_{2}$} & {\scriptsize{}$3.24e^{-4}[\mathrm{kgm^{2}}]$} & {\scriptsize{}$r_{a}$} & {\scriptsize{}$1.9[\mathrm{\Omega}]$}\tabularnewline
\hline 
{\scriptsize{}$J_{3}$} & {\scriptsize{}$4.69e^{-4}[\mathrm{kgm^{2}}]$} & {\scriptsize{}$r_{e}$} & {\scriptsize{}$58$}\tabularnewline
\hline 
{\scriptsize{}$L$} & {\scriptsize{}$0.1100[\mathrm{m}]$} & {\scriptsize{}$\bar{f}_{v}$} & {\scriptsize{}$1e^{-4}I$}\tabularnewline
\hline 
{\scriptsize{}$l_{1}$} & {\scriptsize{}$0.1524[\mathrm{m}]$} & {\scriptsize{}$\bar{f}_{d}$} & {\scriptsize{}$1e^{-4}I$}\tabularnewline
\hline 
{\scriptsize{}$l_{2}$} & {\scriptsize{}$0.1505[\mathrm{m}]$} & {\scriptsize{}--} & {\scriptsize{}--}\tabularnewline
\hline 
\end{tabular}{\scriptsize{}} \label{table:parameters}}
\end{table}

Using some numerical calculations, it is possible to obtain that $g_m^{\Delta}=0.0714$, $g_M^{\Delta}=5.2696$, $\delta_1=2$, $\delta_2=4.0080$, $\delta_3=1.3434$, $\gamma_1^{\Delta}=1.9957$, $\gamma_2^{\Delta}=3.2688$, $\gamma_3=18.4838$, $\gamma_4=2.5907e^{-4}$ and $\gamma_5=4.6706e^{-6}$. Therefore, it is easy to verify that condition \eqref{eq:FeasibCond}, in Theorem \ref{T:gains}, is satisfied; and thus, there exist positive gains $k_{1}$ and $k_{2}$ such that $[s^T,z^T]^{T}=0$ is globally robust finite-time stable. 

It is worth mentioning that one could compute the controller gains following Algorithm \ref{A:gs}. However, due to the fact that such an Algorithm is based on conservative estimations given in \eqref{eq:DefMus} and \eqref{eq:DefThetas}, it is natural that the obtained gains are also conservative, as in this particular example, where very large values are computed.

\begin{figure}[tbh]
\center
\includegraphics[width=8.5cm,height=5cm]{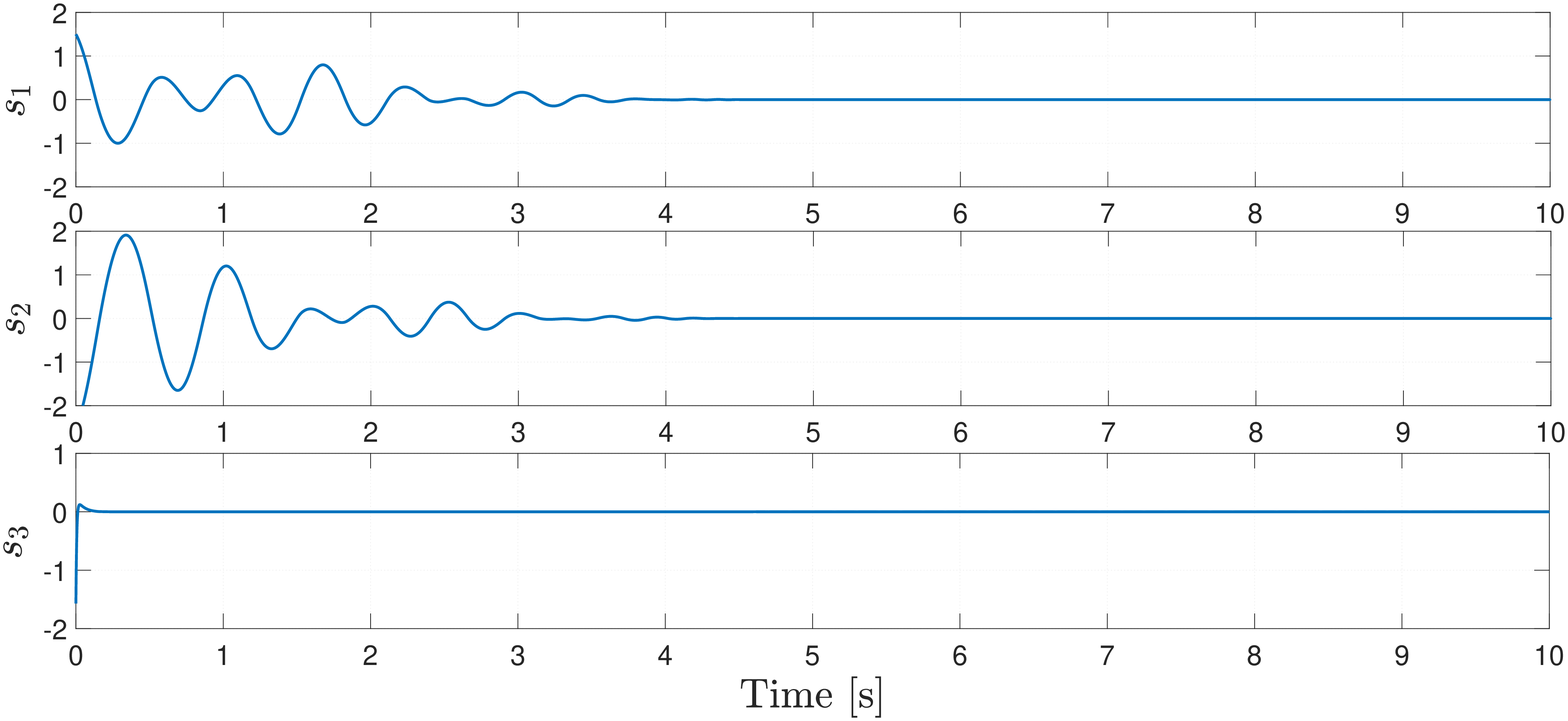} 
\caption{Sliding Variables}
\label{fig:s}
\end{figure}

\begin{figure}[tbh]
\center
\includegraphics[width=8.5cm,height=5cm]{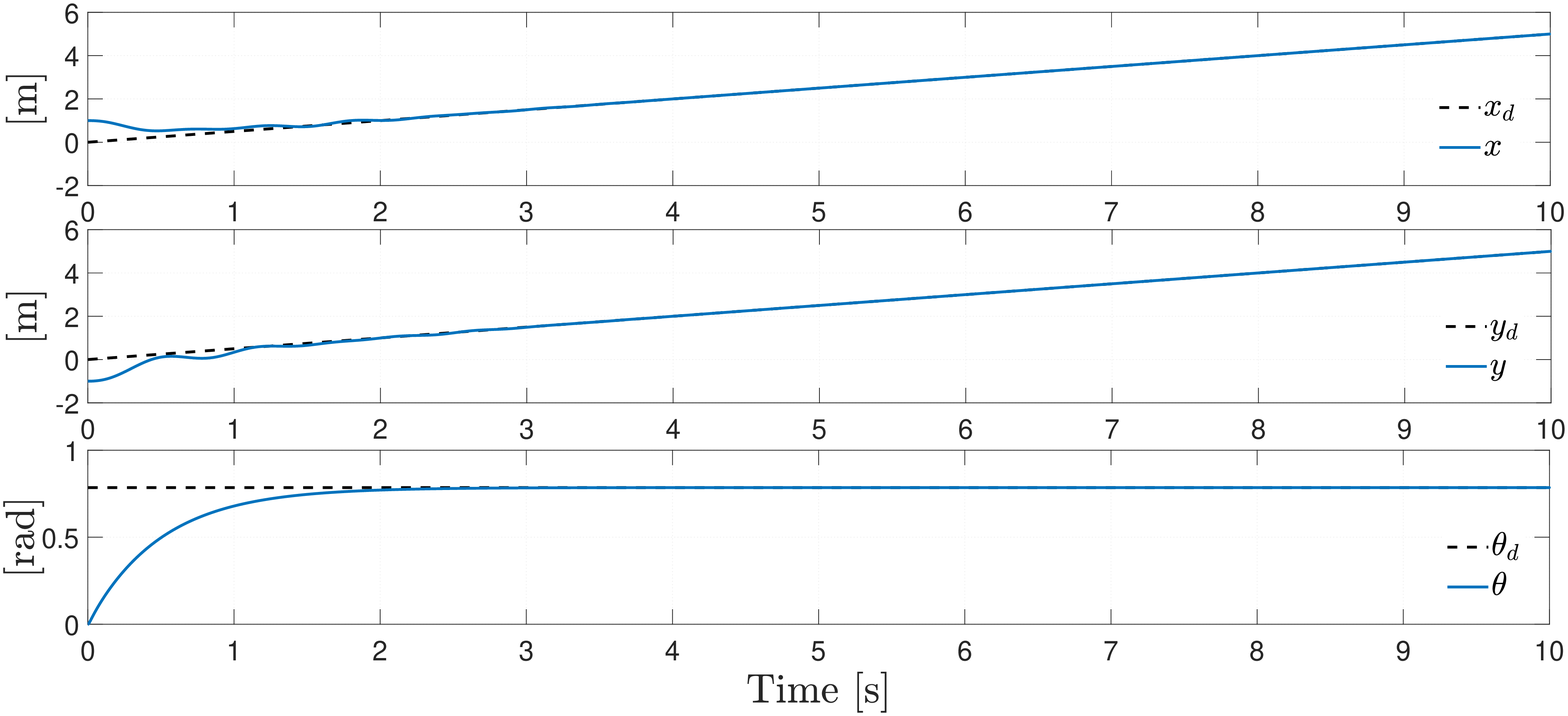} \caption{Configuration Variables}
\label{fig:q}
\end{figure}

\begin{figure}[tbh]
\center
\includegraphics[width=8.5cm,height=5cm]{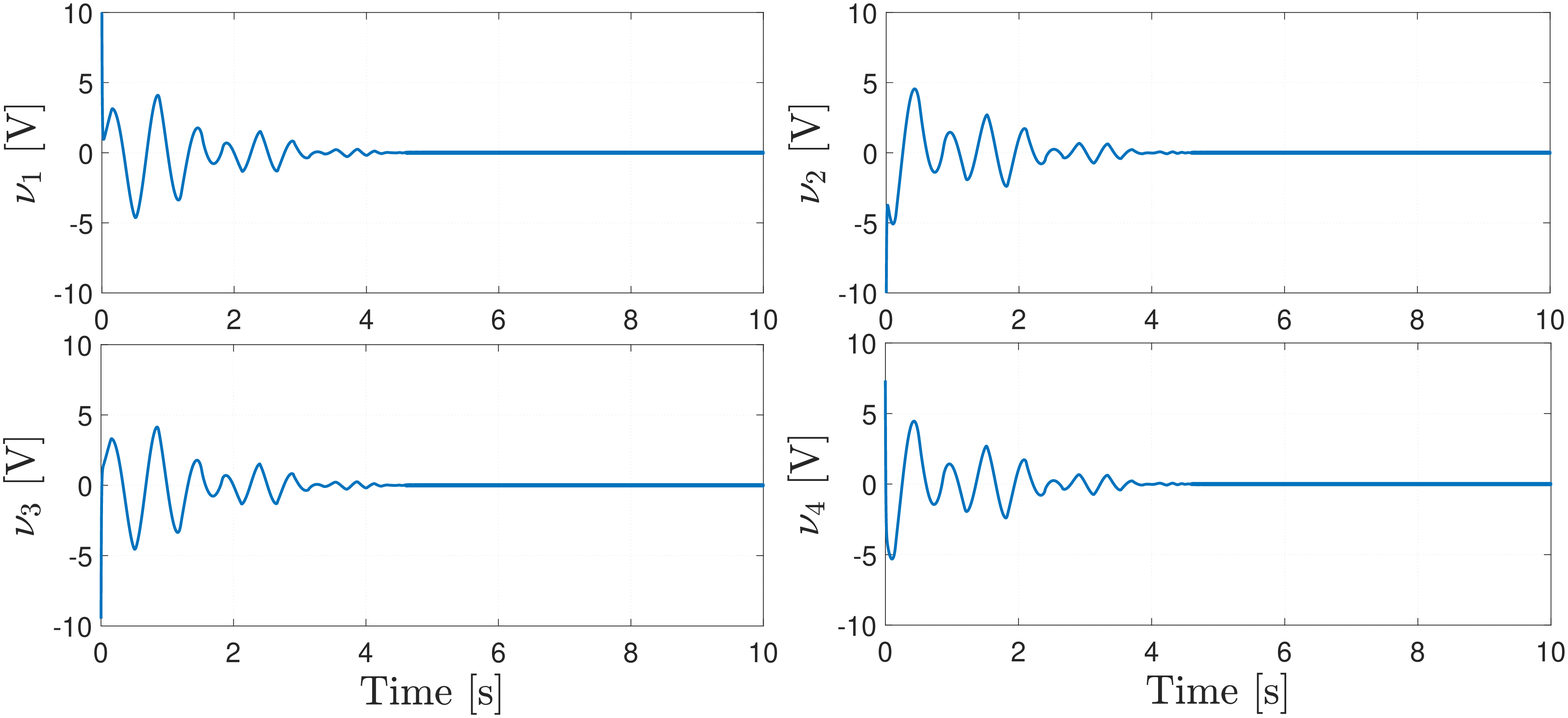} \caption{Armature Voltages}
\label{fig:u}
\end{figure}

For these simulation results, the controller gains are chosen as $k_{1}=42$ and $k_{2}=13$, while $\alpha=1$, $\beta=1$, $b=3$ $p=0.4$ and $\Theta=2I$, ensuring finite-time convergence for $s$. The desired trajectory has been chosen as $q_{d}(t)=[0.5t,0.5t,\pi/4]^{T}$. The obtained results are illustrated by Figs. \ref{fig:s}, \ref{fig:q} and \ref{fig:u}. 

Fig. \ref{fig:s} shows the behavior of the sliding variables. Note
that such variables vanish in a finite time, which illustrates the
main result given in Theorem \ref{T:gains}.

Fig. \ref{fig:q} depicts the configuration variables of the omnidirectional
mobile robot. It can be seen that the mobile robot exponentially tracks
the desired reference even when there exists state-dependent uncertainty
in the input matrix and unknown friction forces. This results highlight
the effectiveness of the proposed approach.

Finally, Fig. \ref{fig:u} shows the of the control inputs. Note that
in this case such inputs are given by the armature voltages, which
are directly applied to each wheel motor through the relation $\nu=E^{+}u$,
with $E^{+}$ the right pseudo-inverse of $E$ and $u$ the current
control signals designed based on the MGSTA. Note that the control
signals do not seem to be excessive for a real omnidirectional mobile
robot. Moreover, it is worth saying that, under ideal conditions,
the MGSTA provides an excellent solution to the chattering problem,
if no actuator dynamics is considered.

\section{Conclusions}
\label{sec:CR} 
This paper presents a Lyapunov approach to the design
of the MGSTA to control systems with perturbations and an uncertain
input matrix, both depending on time and system states. The proposed
procedure provides sufficient conditions for the MGSTA gains selection ensuring global finite-time stability of the system's outputs. The simulation
results illustrate the effectiveness of the proposed version of the MGSTA
for a four-wheeled omnidirectional mobile robot even when the conditions of \cite{edwards},
\cite{vidal1} and \cite{vidal2} cannot be satisfied.

Future research can be devoted to obtain less conservative estimations for the uncertainties of the system and provide better algorithms for the controller gains design. 

\appendix

\section*{Proof of the Main Result}
In this section, following usual ideas for the Super-Twisting \cite{moreno2008,moreno_osorio,moreno,edwards,vidal1,vidal2} and \cite{MorOso12},
a quadratic non-smooth Lyapunov function is used to prove the main
result Theorem \ref{T:gains}, from which Algorithm \ref{A:gs} is
derived.

\textit{Lyapunov Function Candidate and its Derivative:} Consider the closed-loop system given by (\ref{eq:m1}) and (\ref{eq:m2}).
The Lyapunov function candidate is the quadratic form
\begin{equation}
V\left(\zeta\right)=\frac{1}{2}\zeta^{T}P\zeta=\frac{1}{2}\zeta^{T}\left[\begin{array}{cc}
p_{1}I & -I\\
-I & p_{2}I
\end{array}\right]\zeta\,,\label{eq:Lyap}
\end{equation}
in the variable $\zeta=[\zeta_1^T,\zeta_2^T]^T=[\xi^T,z^T]^T$, where $\xi=\phi_{1}(x)$ and $z=v+b^{-1}\left(I+\Delta_{G}\right)^{-1}f_{2}$.
If the positive \emph{constants} $p_{1},p_{2}>0$ are such that $p_{1}p_{2}>1$,
then the function $V\left(\zeta\right)$ is positive definite, radially
unbounded and differentiable almost everywhere, \textit{i.e.,} it is not differentiable
whenever $\xi=x=0$.

To calculate its derivative, we require $\dot{\zeta}$. From the definition
of $J(x)$ in (\ref{eq:J}) it follows that $\dot{\xi}=J(x)\dot{x}$. Using the shorthand notations
\[
\tilde{K}_{1}=k_{1}I-G^{-1}\Delta_{1}\,,\,\tilde{K}_{2}=k_{2}I-\Delta_{2}\,,\,A=\frac{\Delta_{1}}{c}\,,
\]
it follows that $\dot{\zeta}$ is given almost everywhere by
\begin{equation}
\dot{\zeta}=S\left(t,x\right)\zeta,\label{eq:CLS}
\end{equation}
with the square matrix $S(t,x)$ given as 
\begin{gather*}
S\left(t,x\right)=\left[\begin{array}{cc}
-JG\tilde{K}_{1} & J\left(I+\Delta_{G}\right)b\\
-c\tilde{K}_{2}-\Delta_{1}G\tilde{K}_{1}\; & \Delta_{1}\left(I+\Delta_{G}\right)b
\end{array}\right].
\end{gather*}

The derivative of the Lyapunov function along the trajectories of
(\ref{eq:m1})-(\ref{eq:m2}), and using the relation $J\zeta_{1}=c\zeta_{1}$
given in (\ref{eq:ST2}), can be written as 
\[
\dot{V}=-c\zeta^{T}Q\zeta=-c\left[\begin{array}{cc}
Q_{11} & Q_{12}\\
Q_{12}^{T} & Q_{22}
\end{array}\right]\zeta\,,
\]
where 
\begin{subequations} \label{eq:DefQ} 
\begin{align}
Q_{11} & ={\rm Sym}\left\{ p_{1}G\tilde{K}_{1}-\left(\tilde{K}_{2}+AG\tilde{K}_{1}\right)\right\} ,\label{eq:DefQ11}\\
Q_{21} & =-\mathcal{J}G\tilde{K}_{1}+p_{2}\left(\tilde{K}_{2}+AG\tilde{K}_{1}\right)\nonumber \\
 & \qquad\qquad+b\left(I+\Delta_{G}^{T}\right)\left(A^{T}-p_{1}I\right),\label{eq:DefQ21}\\
Q_{22} & =b{\rm Sym}\left\{ \mathcal{J}\left(I+\Delta_{G}\right)-p_{2}A\left(I+\Delta_{G}\right)\right\}.
\end{align}
\end{subequations}

\textit{Positive Definiteness of $Q$:} The negative definiteness of $\dot{V}$ and the global finite-time
stability of the origin $x=0,\,z=0$ follows from usual arguments
for the Super-Twisting (see, \textit{e.g.}, \cite{moreno} and \cite{MorOso12}) if the bounded matrix
$Q\left(t,\,x\right)$ is positive definite for all values of $\left(t,\,x\right)$.
For $Q$ to be positive definite it is necessary that $Q_{22}$ is
also positive definite. Using (\ref{eq:DefMus}) and assuming that
$p_{2}>0$ is chosen sufficiently small so that (\ref{eq:Cond1p2})
is satisfied, it is readily obtained that 
\begin{equation}
0<b\tilde{\gamma}_{1}^{\Delta}I\leq Q_{22},\label{eq:IneqQ22}
\end{equation}
where $\tilde{\gamma}_{1}^{\Delta}=\gamma_{1}^{\Delta}-p_{2}\mu_{4}$,
showing that $Q_{22}$ is positive definite. Moreover, $Q$ is positive
definite if so is the Schur Complement, \textit{i.e.,} 
\begin{equation}
Q_{11}>Q_{21}^{T}Q_{22}^{-1}Q_{21}.\label{eq:InvSchr}
\end{equation}

Note that 
\[
Q_{11}={\rm Sym}\left\{ k_{1}p_{1}G-k_{2}I-p_{1}\Delta_{1}-k_{1}AG+\Delta_{2}+A\Delta_{1}\right\} ,
\]
so that the use of (\ref{eq:Gbounds}) and (\ref{eq:DefMus}) in (\ref{eq:DefQ11})
leads to 
\begin{equation}
Q_{11}>g_{m}^{\Delta}p_{1}k_{1}I-2k_{2}I-\left(p_{1}\mu_{1}+k_{1}\mu_{2}+\mu_{3}\right)I.\label{eq:IneqQ11}
\end{equation}

From (\ref{eq:IneqQ22}) it easily follows that $Q_{21}^{T}Q_{22}^{-1}Q_{21}\leq\frac{1}{b\tilde{\gamma}_{1}^{\Delta}}Q_{21}^{T}Q_{21}$.
Then, selecting $k_{2}$ as in (\ref{eq:Selectionk2}), in order to
cancel some known terms in $Q_{21}$, from its expression in (\ref{eq:DefQ21})
and with the help of (\ref{eq:DefThetas}), after a simple but lengthy
calculation, the following inequality is attained 
\begin{multline}
Q_{21}^{T}Q_{22}^{-1}Q_{21}\leq\frac{1}{b\tilde{\gamma}_{1}^{\Delta}}(\gamma_{3}k_{1}^{2}+\theta_{1}p_{2}k_{1}^{2}+\theta_{2}p_{2}^{2}k_{1}^{2}+\theta_{3}k_{1}\\
+\gamma_{4}bp_{1}k_{1}+\theta_{4}p_{2}k_{1}+\theta_{5}p_{2}k_{1}+\theta_{6}bp_{1}p_{2}k_{1}\\
\theta_{7}p_{2}^{2}k_{1}+\gamma_{5}b^{2}p_{1}^{2}+\theta_{8}bp_{1}+\theta_{9}bp_{1}p_{2}\\
+\theta_{10}p_{2}+\theta_{11}p_{2}^{2}+\theta_{12})I.\label{eq:IneqRHS}
\end{multline}

Using (\ref{eq:IneqQ11}) and (\ref{eq:IneqRHS}) in (\ref{eq:InvSchr}),
and rearranging the terms as a polynomial in $k_{1}$, it is concluded
that the inequality 
\begin{equation}
\alpha_{2}k_{1}^{2}+\alpha_{1}k_{1}+\alpha_{0}<0\,,\label{eq:Polyk1}
\end{equation}
where $\alpha_{i}$, with $i=0,\,1,\,2$, are defined in (\ref{eq:DefAlphas}),
implies the positive definiteness of $Q$. The existence of positive values of the gain $k_{1}$ such that inequality
(\ref{eq:Polyk1}) is fulfilled requires the polynomial to have two
positive real roots (since $\alpha_{0}>0$), \textit{i.e.}, $\alpha_{1}^{2}-4\alpha_{2}\alpha_{0}>0$ and $\alpha_{1}<0.$ These two inequalities are equivalent to (\ref{eq:IneqPModb}) and
(\ref{eq:IneqPMod}). To conclude the proof, we need to show that the main condition (\ref{eq:FeasibCond}) implies (\ref{eq:Cond1p2})-(\ref{eq:IneqPMod}), which are the inequalities
used up to this point. In order to show that, fix $b>0$ and note
that (\ref{eq:IneqPModb}) is $\gamma_{1}^{\Delta}g_{m}^{\Delta}-\gamma_{4}-\Gamma_{0}\left(p_{1},\,p_{2}\right)>0$,
with $\Gamma_{0}$ a continuous function, and $\lim_{p_{1}\rightarrow0}\left\{ \lim_{p_{2}\rightarrow\infty}\Gamma_{0}\left(p_{1},p_{2}\right)\right\} =0$.
As a consequence, (\ref{eq:FeasibCond}) implies that (\ref{eq:IneqPModb})
is satisfied for values of $\left(p_{1},p_{2}\right)$ with $p_{1}$
large and $p_{2}$ small. Note that $p_{2}$ sufficiently small also implies
(\ref{eq:Cond1p2}).

A similar reasoning shows that $\Gamma_{1}\left(p_{1},p_{2}\right)$,
the coefficient of $p_{1}$ in (\ref{eq:IneqPMod}), is positive for
large values of $p_{1}$ and small values of $p_{2}$. Fixing the
value of $p_{2}$, it is easy to see that $\Gamma_{1}\left(p_{1},p_{2}\right)p_{1}$
in (\ref{eq:IneqPMod}) grows unboundedly and strictly monotonically
with $p_{1}$ (for $p_{1}$ sufficiently large), while $\Gamma_{2}\left(p_{1},p_{2}\right)$
is bounded. This implies that, for all sufficiently large values of
$p_{1}$, (\ref{eq:IneqPMod}) will be satisfied. Moreover, $p_{1}$
can be selected sufficiently large so that $p_{1}p_{2}>1$, and thus
$V>0$. Therefore, there exist appropriate values of $k_{1}$ to render
$Q>0$, \textit{i.e.}, $\dot{V}<0$, so that the system is asymptotically
stable. Due to the dominance of the negative homogeneity degree, finite-time
convergence is obtained. This concludes the proof.

\bibliographystyle{ieeetr}
\bibliography{biblio}

\end{document}